\documentclass{article}
\usepackage{ismir,cite,url}
\usepackage{amsmath,amsfonts,amssymb}
\usepackage{graphicx}
\usepackage{color}
\usepackage{booktabs, multirow}
\usepackage{microtype}
\usepackage{tabulary}

\newcommand\blfootnote[1]{%
  \begingroup
  \renewcommand\thefootnote{}\footnote{#1}%
  \addtocounter{footnote}{-1}%
  \endgroup
}

\title{Wave-U-Net: A Multi-Scale Neural Network for End-to-End Audio Source Separation}  

\threeauthors
  {Daniel Stoller} {Queen Mary University of London \\ {\tt d.stoller@qmul.ac.uk}}
  {Sebastian Ewert} {Spotify\\ {\tt sewert@spotify.com}}
  {Simon Dixon} {Queen Mary University of London \\ {\tt s.e.dixon@qmul.ac.uk}}

\begin{document}

\maketitle

\renewcommand{\baselinestretch}{0.97}
\selectfont

\begin{abstract}
Models for audio source separation usually operate on the magnitude spectrum, which ignores phase information and makes separation performance dependant on hyper-parameters for the spectral front-end.
Therefore, we investigate end-to-end source separation in the time-domain, which allows modelling phase information and avoids fixed spectral transformations.
Due to high sampling rates for audio, employing a long temporal input context on the sample level is difficult, but required for high quality separation results because of long-range temporal correlations.
In this context, we propose the Wave-U-Net, an adaptation of the U-Net to the one-dimensional time domain, which repeatedly resamples feature maps to compute and combine features at different time scales.
We introduce further architectural improvements, including an output layer that enforces source additivity, an upsampling technique and a context-aware prediction framework to reduce output artifacts.
Experiments for singing voice separation indicate that our architecture yields a performance comparable to a state-of-the-art spectrogram-based U-Net architecture, given the same data.
Finally, we reveal a problem with outliers in the currently used SDR evaluation metrics and suggest reporting rank-based statistics to alleviate this problem.
\end{abstract}

\section{Introduction}
\label{sec:intro}

Current methods for audio source separation almost exclusively operate on spectrogram representations of the audio signals~\cite{Huang2014,Jansson2017}, as they allow for direct access to components in time and frequency.
In particular, after applying a short-time Fourier transform (STFT) to the input mixture signal, the complex-valued spectrogram is split into its magnitude and phase components.
Then only the magnitudes are input to a parametric model, which returns estimated spectrogram magnitudes for the individual sound sources.
To generate corresponding audio signals, these magnitudes are combined with the mixture phase and then converted with an inverse STFT to the time domain.
Optionally, the phase can be recovered for each source individually using the Griffin-Lim algorithm~\cite{Griffin1984}.

This approach has several limitations.
Firstly, the STFT output depends on many parameters, such as the size and overlap of audio frames, which can affect the time and frequency resolution.
Ideally, these parameters should be optimised in conjunction with the parameters of the separation model to maximise performance for a particular separation task.
In practice, however, the transform parameters are fixed to specific values.
Secondly, since the separation model does not estimate the source phase, it is often assumed to be equal to the mixture phase, which is incorrect for overlapping partials.
Alternatively, the Griffin-Lim algorithm can be applied to find an approximation to a signal whose magnitudes are equal to the estimated ones, but this is slow and often no such signal exists~\cite{LeRoux2008}.
Lastly, the mixture phase is ignored in the estimation of sources, which can potentially limit the performance.
Thus, it would be desirable for the separation model to learn to estimate the source signals including their phase directly.

\blfootnote{This work was partially funded by EPSRC grant EP/L01632X/1. Implementation available at \url{https://github.com/f90/Wave-U-Net}}

As an approach to tackle the above problems, several audio processing models were recently proposed that operate directly on time-domain audio signals, including speech denoising as a task related to general audio source separation~\cite{Dieleman2016,Pascual2017,Rethage2017}.
Inspired by these first results, we investigate in this paper the potential of fully end-to-end time-domain separation systems in the face of unresolved challenges.
In particular, it is not clear if such a system will be able to deal effectively with the very long-range temporal dependencies present in audio due to its high sampling rate.
Further, it is not obvious upfront whether the additional phase information will indeed be beneficial for the task, or whether the noisy phase might be detrimental for the learning dynamics in such a system.
Overall, our contributions in this paper can be summarised as follows.
\begin{itemize}
\item{We propose the Wave-U-Net, a one-dimensional adaptation of the U-Net architecture~\cite{Ronneberger2015,Jansson2017}, which separates sources directly in the time domain and can take large temporal contexts into account.}
\item{We show a way to provide the model with additional input context to avoid artifacts at the boundaries of output windows, in contrast to previous work~\cite{Pascual2017,Jansson2017}.}
\item{We replace strided transposed convolution used in previous work~\cite{Jansson2017,Pascual2017} for upsampling feature maps with linear interpolation followed by a normal convolution to avoid artifacts.}
\item{The Wave-U-Net achieves good multi-instrument and singing voice separation, the latter of which compares favourably to our re-implementation of the state-of-the-art network architecture \cite{Jansson2017}, which we train under comparable settings.}
\item{Since the Wave-U-Net can process multi-channel audio, we compare stereo with mono source separation performance}
\item{We highlight an issue with the commonly used Signal-to-Distortion ratio evaluation metric, and propose a work-around.}
\end{itemize}

It should be noted that we expect the current state of the art model as presented in \cite{Jansson2017} to yield higher separation quality than what we report here, as the training dataset used in \cite{Jansson2017} is well-designed, highly unbiased and considerably larger. However, we believe that our comparison with a re-implementation trained under similar conditions might be indicative of relative performance improvements. 

\begin{figure}[t]
\centering
\centerline{\includegraphics[width=8.5cm]{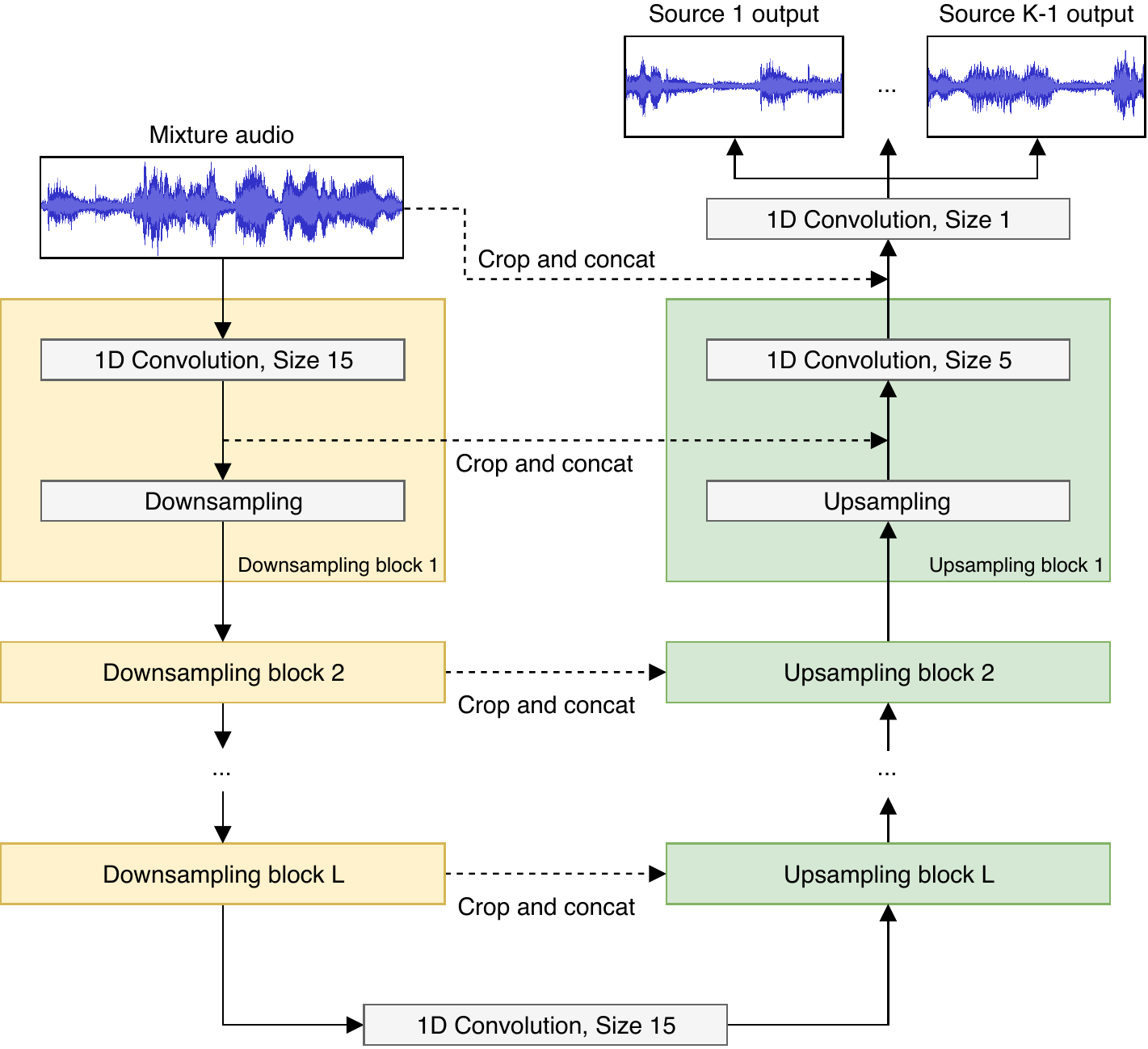}}
\caption{Our proposed Wave-U-Net with $K$ sources and $L$ layers. With our difference output layer, the $K$-th source prediction is the difference between the mixture and the sum of the other sources.}
\label{fig:waveunet}
\end{figure}

\section{Related work}
\label{sec:related_work}

To alleviate the problem of fixed spectral representations widely used in previous work~\cite{Nugraha2015,Simpson2015, Miron2017, Huang2014,Luo2017,Uhlich2017}, an adaptive front-end for spectrogram computation was developed~\cite{Venkataramani2017} that is trained jointly with the separation network, which operates on the resulting magnitude spectrogram.
Despite comparatively increased performance, the model does not exploit the mixture phase for better source magnitude predictions and also does not output the source phase, so the mixture phase has to be used for source signal reconstruction, both of which limit performance.

To our knowledge, only the TasNet~\cite{Luo2017a} and MRCAE~\cite{Grais2018} systems tackle the general problem of audio source separation in the time domain.
The TasNet performs a decomposition of the signal into a set of basis signals and weights, and then creates a mask over the weights which are finally used to reconstruct the source signals. The model is shown to work for a speech separation task.
However, the work makes conceptual trade-offs to allow for low-latency applications, while we focus on offline application, allowing us to exploit a large amount of contextual information. 

The multi-resolution convolutional auto-encoder (MRCAE)~\cite{Grais2018} uses two layers of convolution and transposed convolution each.
The authors argue the different convolutional filter sizes detect audio frequencies with different resolutions, but they work only on one time resolution (that of the input), since the network does not perform any resampling.
Since input and output consist of only 1025 audio samples (equivalent to 23 ms), it can only exploit very little context information.
Furthermore, at test time, output segments are overlapped using a regular spacing and then combined, which differs from how the network is trained. This mismatch and the small context could hurt performance and also explain why the provided sound examples exhibit many artifacts.

For the purpose of speech enhancement and denoising, the SEGAN~\cite{Pascual2017} was developed, employing a neural network with an encoder and decoder pathway that successively halves and doubles the resolution of feature maps in each layer, respectively, and features skip connections between encoder and decoder layers. While we use a similar architecture, we rectify the issue of aliasing artifacts in the generated output when using strided transposed convolutions as shown by~\cite{Odena2016}.
Furthermore, the model cannot predict audio samples close to its border output well since it is given no additional input context, which is an issue we address using convolutions with proper padding.
It is also not clear if the model's performance can transfer to other and more challenging audio source separation tasks.

The Wavenet~\cite{Dieleman2016} was adapted for speech denoising~\cite{Rethage2017} to have a non-causal conditional input and a parallel output of samples for each prediction and is based on the repeated application of dilated convolutions with exponentially increasing dilation factors to factor in context information.
While this architecture is very parameter-efficient, memory consumption is high since each feature map resulting from a dilated convolution still has the original audio's sampling rate as resolution.

In contrast, our approach calculates the longer-term dependencies based on feature maps with more features and increasingly lower resolution. This saves memory and enables a large number of high-level features, which arguably do not need sample-level resolution to be useful, such as instrument activity, or the position in the current measure.

\section{The Wave-U-Net model}
\label{sec:model}

Our goal is to separate a mixture waveform ${\mathbf{M} \in [-1,1]^{L_m \times\ C}}$ into $K$ source waveforms $\mathbf{S}^1, \ldots, \mathbf{S}^K$ with ${\mathbf{S}^k \in [-1,1]^{L_s \times\ C}}$ for all ${k \in \{1, \ldots, K\}}$, $C$ as the number of audio channels and $L_m$ and $L_s$ as the respective numbers of audio samples.
For model variants with extra input context, we have $L_m > L_s$ and make predictions for the centre part of the input.

\subsection{The base architecture}
\label{sec:model_base}

\begin{table}[t]
\centering
\begingroup
\footnotesize
\begin{tabulary}{0.48\textwidth}{|C|CC|}
\hline
Block & Operation & Shape \\
\hline \hline
& Input & $(16384, 1)$ \\
\hline
\multirow{2}*{\begin{tabular}{c} DS, repeated for \\ $i=1,\ldots,L$ \\
                \end{tabular}} & \texttt{Conv1D}($F_c \cdot i$, $f_d$) & \\
& \texttt{Decimate} & $(4, 288)$ \\
\hline
& \texttt{Conv1D}($F_c \cdot (L+1)$, $f_d$) & $(4, 312)$ \\
\hline
\multirow{3}*{\begin{tabular}{c} US, repeated for \\ $i=L,\ldots,1$ \\
                \end{tabular}} & \texttt{Upsample} & \\
& \texttt{Concat}(DS block $i$) & \\
& \texttt{Conv1D}($F_c \cdot i$, $f_u$) & $(16834, 24)$ \\
\hline
& \texttt{Concat}(Input) & $(16834, 25)$ \\
& \texttt{Conv1D}($K$, 1) & $(16834, 2)$ \\
\hline
\end{tabulary}
\endgroup
\caption{Block diagram of the base architecture. Shapes describe the final output after potential repeated application of blocks, for the example of model M1, and denote the number of time steps and feature channels, in that order. DS block $i$ refers to the output before decimation. Note that the US blocks are applied in reverse order, from level $L$ to 1.}
\label{tab:blocks}
\end{table}

A diagram of the Wave-U-Net architecture is shown in Figure~\ref{fig:waveunet}.
It computes an increasing number of higher-level features on coarser time scales using downsampling (DS) blocks.
These features are combined with the earlier computed local, high-resolution features using upsampling (US) blocks, yielding multi-scale features which are used for making predictions.
The network has $L$ levels in total, with each successive level operating at half the time resolution as the previous one.
For $K$ sources to be estimated, the model returns predictions in the interval $(-1,1)$, one for each source audio sample.

The detailed architecture is shown in Table~\ref{tab:blocks}.
\texttt{Conv1D}(x,y) denotes a 1D convolution with $x$ filters of size $y$.
It includes zero-padding for the base architecture, and is followed by a LeakyReLU activation (except for the final one, which uses $\tanh$).
\texttt{Decimate} discards features for every other time step to halve the time resolution.
\texttt{Upsample} performs upsampling in the time direction by a factor of two, for which we use linear interpolation (see Section~\ref{sec:model_artifacts} for details).
\texttt{Concat}(x) concatenates the current, high-level features with more local features x.
In extensions of the base architecture (see below), where \texttt{Conv1D} does not involve zero-padding, x is centre-cropped first so it has the same number of time steps as the current layer.

\subsubsection{Avoiding aliasing artifacts due to upsampling}
\label{sec:model_artifacts}

\begin{figure}[t]
\centering
\centerline{\includegraphics[width=8.0cm]{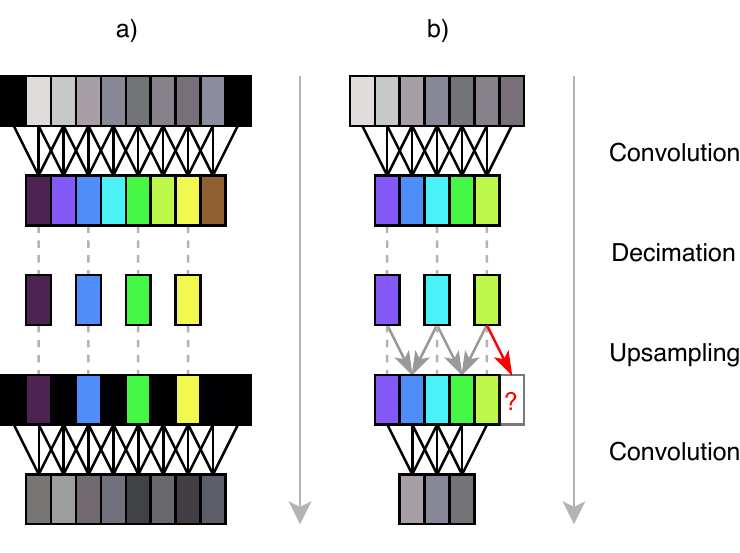}}
\caption{a) Common model (e.g.~\cite{Jansson2017}) with an even number of inputs (grey) which are zero-padded (black) before convolving, creating artifacts at the borders (dark colours). After decimation, a transposed convolution with stride 2 is shown here as upsampling by zero-padding intermediate and border values followed by normal convolution, which likely creates high-frequency artifacts in the output. b) Our model with proper input context and linear interpolation for upsampling from Section~\ref{sec:model_improv_context} does not use zero-padding. The number of features is kept uneven, so that upsampling does not require extrapolating values (red arrow). Although the output is smaller, artifacts are avoided.}
\label{fig:convpaddingupsampling}
\end{figure}

Many related approaches use transposed convolutions with strides to upsample feature maps~\cite{Pascual2017,Jansson2017}.
This can introduce aliasing effects in the output, as shown for the case of image generation networks~\cite{Odena2016}.
In initial tests, we also found artifacts when using such convolutions as upsampling blocks in our Wave-U-Net model in the form of high-frequency buzzing noise.

Transposed convolutions with a filter size of $k$ and a stride of $x > 1$ can be viewed as convolutions applied to feature maps padded with $x-1$ zeros between each original value~\cite{Dumoulin2016a}.
We suspect that the interleaving with zeros without subsequent low-pass filtering introduces high-frequency patterns into the feature maps, shown symbolically in Figure~\ref{fig:convpaddingupsampling}, which leads to high-frequency noise in the final output as well.
Instead of transposed strided convolutions, we thus perform linear interpolation for upsampling, which ensures temporal continuity in the feature space, followed by a normal convolution.
In initial tests, we did not observe any high-frequency sound artifacts in the output with this technique and achieved very similar performance.

\subsection{Architectural improvements}
\label{sec:model_improv}

The previous Section described the baseline variant of the Wave-U-Net.
In the following, we will describe a set of architectural improvements for the Wave-U-Net designed to increase model performance.

\subsubsection{Difference output layer}
\label{sec:model_improv_diff}

Our baseline model outputs one source estimate for each of $K$ sources by independently applying $K$ convolutional filters followed by a $\tanh$ non-linearity to the last feature map.
In the separation tasks we consider, the mixture signal is the sum of its source signal components: $\mathbf{M} \approx \sum_{j=1}^K \mathbf{S}^j$. 
Since our baseline model is not constrained in this fashion, it has to learn this rule approximately to avoid highly improbable outputs, which could slow down learning and reduce performance.
Therefore, we use a difference output layer to constrain the outputs $\hat{\mathbf{S}}^j$, enforcing $\sum_{j=1}^K \hat{\mathbf{S}}^j = \mathbf{M}$:
only $K-1$ convolutional filters with a size of 1 are applied to the last feature map of the network, followed by a $\tanh$ non-linearity, to estimate the first $K-1$ source signals.
The last source is then simply computed as $\hat{\mathbf{S}}^K=\mathbf{M} - \sum_{j=1}^{K-1} \hat{\mathbf{S}}^j$.

This type of output was also used for speech denoising in~\cite{Rethage2017} as part of an ``energy-conserving" loss, and a similar idea can be found very commonly in spectrogram-based source separation in the form of masks that distribute the energy of the input mixture magnitudes to the output sources.
We investigate the impact of introducing this layer and its additivity assumption, since it depends on the extent to which this additivity property is satisfied by the data.

\subsubsection{Prediction with proper input context and resampling}
\label{sec:model_improv_context}

In previous work~\cite{Jansson2017,Grais2018,Pascual2017}, the input and the feature maps are padded with zeros before convolving, so that the resulting feature map does not change in its dimension, as shown in Figure~\ref{fig:convpaddingupsampling}a.
This simplifies the network's implementation, since the input and output dimensions are the same.
Zero-padding audio or spectrogram input this way effectively extends the input using silence at the beginning and end.
However, taken from a random position in a full audio signal,
the information at the boundary becomes artificial, i.e.\ the temporal context for this excerpt is given in the full audio signal but is ignored and assumed to be silent.
Without proper context information, the network thus has difficulty predicting output values near the beginning and end of the sequence.
As a result, simply concatenating the outputs as non-overlapping segments at test time to obtain the prediction for a full audio signal can create audible artifacts at the segment borders, as neighbouring outputs can be inconsistent when they are generated without correct context information.
In Section~\ref{sec:results_qualitative}, we investigate this behaviour in practice.

As a solution, we employ convolutions without implicit padding and instead provide a mixture input larger than the size of the output prediction, so that the convolutions are computed on the correct audio context (see Figure~\ref{fig:convpaddingupsampling}b).
Since this reduces the feature map sizes, we constrain the possible output sizes of the network so that feature maps are always large enough for the following convolution.

Further, when resampling feature maps, feature dimensions are often exactly halved or doubled~\cite{Jansson2017,Pascual2017}, as shown in Figure~\ref{fig:convpaddingupsampling}a for transposed strided convolution.
However, this necessarily involves extrapolating at least one value at a border, which can again introduce artifacts.
Instead, we interpolate only between known neighbouring values and keep the very first and last entries, producing $2n-1$ entries from $n$ or vice versa, as shown in Figure~\ref{fig:convpaddingupsampling}b.
To recover the intermediate values after decimation, while keeping border values the same, we ensure that feature maps have odd dimensionality.

\subsubsection{Stereo channels}
\label{sec:model_improv_stereo}

To accommodate for multi-channel input with $C$ channels, we simply change the input $\mathbf{M}$ from an $L_m \times 1$ to an $L_m \times C$ matrix. Since the second dimension is treated as a feature channel, the first convolution of the network takes into account all input channels.
For multi-channel output with $C$ channels, we modify the output component to have $K$ independent convolutional layers with filter size 1 and $C$ filters each.
With a difference output layer, we only use $K-1$ such convolutional layers.
We use this simple approach with $C=2$ to perform experiments with stereo recordings and investigate the degree of improvement in source separation metrics when using stereo instead of mono estimation.

\subsubsection{Learned upsampling for Wave-U-Net}
\label{sec:model_improv_upsampling}

Linear interpolation for upsampling is simple, parameterless and encourages feature continuity.
However, it may be restricting the network capacity too much.
Perhaps, the feature spaces used in these feature maps are not structured so that a linear interpolation between two points in feature space is a useful point on its own, so that a learned upsampling could further enhance performance.
To this end, we propose the learned upsampling layer.
For a given $F \times n$ feature map with $n$ time steps, we compute an interpolated feature $f_{t+0.5} \in \mathbb{R}^F$ for pairs of neighbouring features $f_t, f_{t+1} \in \mathbb{R}^F$ using parameters $w \in \mathbb{R}^F$ and the sigmoid function $\sigma$ to constrain each $w_i \in w$ to the $[0,1]$ interval:
\begin{equation}
f_{t+0.5} = \sigma(w) \odot f_{t} + (1 - \sigma(w)) \odot f_{t+1}
\end{equation}
This can be implemented as a 1D convolution across time with $F$ filters of size two and no padding with a properly constrained matrix.
The learned interpolation layer can be viewed as a generalisation of simple linear interpolation, since it allows convex combinations of features with weights other than $0.5$.

\section{Experiments}
\label{sec:experiment}

We evaluate the performance of our models on two tasks: Singing voice separation and music separation with bass, drums, guitar, vocals and ``other" instruments as categories, as defined by the SiSec separation campaign~\cite{Liutkus2017}.

\subsection{Datasets}

$75$ tracks from the training partition of the MUSDB~\cite{Rafii2017} multi-track database are randomly assigned to our training set, and the remaining $25$ tracks form the validation set, which is used for early stopping.
Final performance is evaluated on the MUSDB test partition comprised of 50 songs.
For singing voice separation, we also add the whole CCMixter database~\cite{Liutkus2015} to the training set.

As data augmentation for both tasks, we multiply source signals with a factor chosen uniformly from the interval $[0.7, 1.0]$ and set the input mixture as the sum of source signals.
No further data preprocessing is performed, only a conversion to mono (except for stereo models) and downsampling to 22050 Hz.

\subsection{Training procedure}

During training, audio excerpts are sampled randomly and inputs padded accordingly for models with input context.
As loss, we use the mean squared error (MSE) over all source output samples in a batch.
We use the ADAM optimizer with learning rate $0.0001$, decay rates $\beta_1=0.9$ and $\beta_2=0.999$ and a batch size of 16.
We define 2000 iterations as one epoch, and perform early stopping after 20 epochs of no improvement on the validation set, measured by the MSE loss.
Afterwards, the last model is fine-tuned further, with the batch size doubled and the learning rate lowered to $0.00001$, again until 20 epochs without improvement in validation loss.
Finally, the model with the best validation loss is selected.

\subsection{Model settings and variants}

For our baseline model, we use $L_m=L_s=16384$ input and output samples, $L=12$ layers, $F_c=24$ extra filters per layer and filter sizes $f_d=15$ and $f_u=5$.

To determine the impact of the model improvements described in Section~\ref{sec:model_improv}, we train a baseline model M1 as described in Section~\ref{sec:model_base} and models M2 to M5 which add the difference output layer from Section~\ref{sec:model_improv_diff} (M2), the input context and resampling from Section~\ref{sec:model_improv_context} (M3), stereo channels from Section~\ref{sec:model_improv_stereo} (M4), and learned upsampling from Section~\ref{sec:model_improv_upsampling} (M5), and also contain all features of the respectively previous model.
We apply the best model of the above (M4) to multi-instrument separation (M6).
Models with input context (M3 to M6) have $L_m=147443$ input and $L_s=16389$ output samples.

For comparison with previous work, we also train the spectrogram-based U-Net architecture~\cite{Jansson2017} (U7) that achieved state-of-the-art vocal separation performance, and a Wave-U-Net comparison model (M7) under the same conditions, both using the audio-based MSE loss and mono signals downsampled to 8192 Hz.
M7 is based on the best model M4, but is set to $L_m=233459$ and $L_s=102405$ to have very similar output size compared to U7 ($L_s=98650$ samples), $F_c=34$ to bring our network to the same size as U7 (20M param.), and the initial batch size is set to four due to the high amount of memory needed per sample.
To train U7, we backpropagate the error through the inverse STFT operation that is used to construct the source audio signal from the estimated spectrogram magnitudes and the mixture phase.
We also train the same model with an L1 loss on the spectral magnitudes (U7a), following~\cite{Jansson2017}.
Since the training procedure and loss are exactly the same for networks U7 and M7, we can fairly compare both architectures by ensuring that performance differences do not arise simply because of the amount of training data or the type of loss function used, and also compare with a spectrogram-based loss (U7a).
Despite our effort to enable an overall model comparison, note that some training settings such as learning rates used in~\cite{Jansson2017} might differ from ours (and are partly unknown) and could provide better performance with U7 and U7a than shown here, even with the same dataset.

\section{Results}
\label{sec:results}

\subsection{Quantitative results}
\label{sec:results_quantitative}

\subsubsection{Evaluation metrics}

The signal-to-distortion (SDR) metric is commonly used to evaluate source separation performance~\cite{Vincent2006}.
An audio track is usually partitioned into non-overlapping audio segments multiple seconds in length, and segment-wise metrics are then averaged over each audio track or the whole dataset to evaluate model performance.
Following the procedure used for the SiSec separation campaign 2018~\cite{Rafii2017}, these segments are one second long.

\subsubsection{Issues with current evaluation metrics}
\label{sec:results_quantitative_issues}

\begin{figure}[t]
\centering
\centerline{\includegraphics[width=8.5cm]{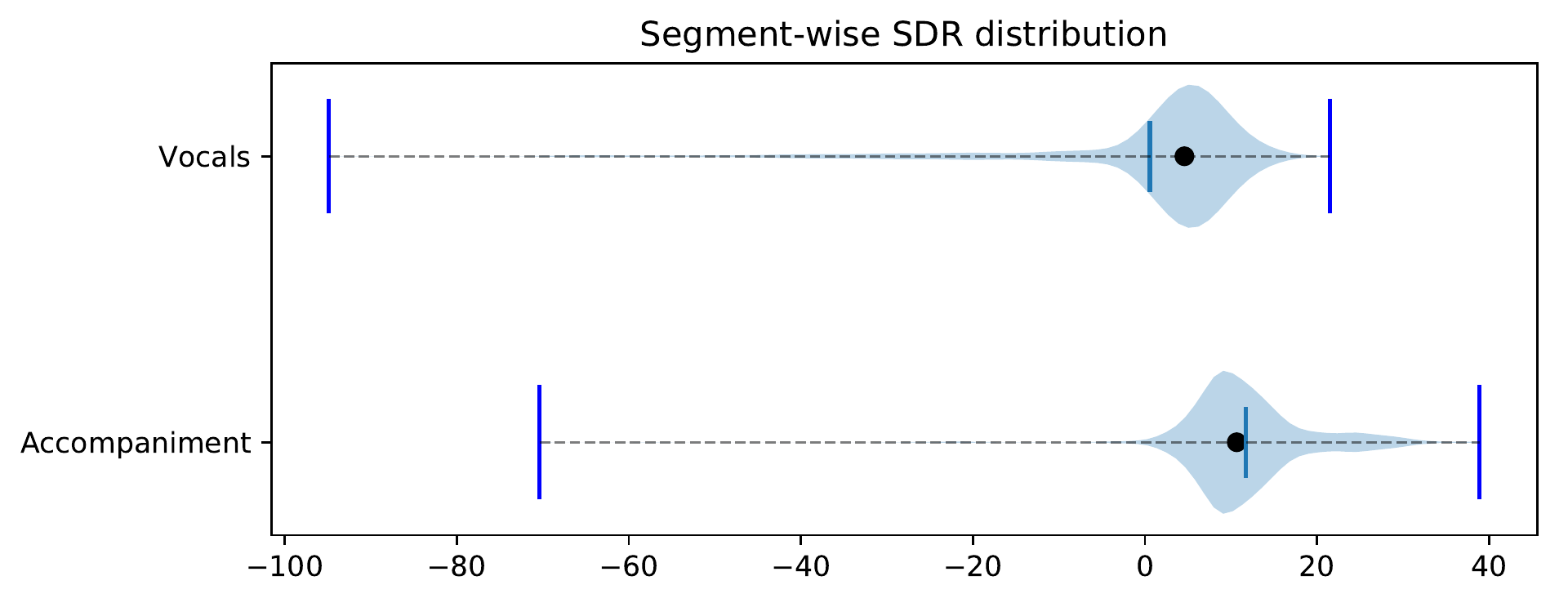}}
\caption{Violin plot of the segment-wise SDR values in the MUSDB test set for model M5. Black points show medians, dark blue lines the means.}
\label{fig:sdr}
\end{figure}

The SDR computation is problematic when the true source is silent or near-silent.
In case of silence, the SDR is undefined ($\log(0)$), which happens often for vocal tracks. Such segments are excluded from the results, so performance on these segments is ignored.
For near-silent parts, the SDR is typically very low when the separator output is quiet, but not silent, although such an output is arguably not a grave error perceptually.
These outliers are visualised using model M5 in Figure~\ref{fig:sdr}.
Since the mean over segments is usually used to obtain overall performance measures, these outliers greatly affect evaluation results.

Since the collection of segment-wise vocal SDR values across the dataset is not normally distributed (compare Figure~\ref{fig:sdr} for vocals), the mean and standard deviation are not sufficient to adequately summarise it.
As a workaround, we take the median over segments, as it is robust against outliers and intuitively describes the minimum performance that is achieved 50\% of the time.
To describe the spread of the distribution, we use the median absolute deviation (MAD) as a rank-based equivalent to the standard deviation (SD).
It is defined as the median of the absolute deviations from the overall median and is easily interpretable, since a value of $x$ means that 50\% of values have an absolute difference from the median that is lower than $x$.

We also note that increasing the duration of segments beyond one second alleviates this issue by removing many, but not all outliers. This is more memory-intensive and presumably still punishes errors during silent sections most.

\subsubsection{Model comparison}
\label{sec:results_comparison}

Table~\ref{tab:results} shows the evaluation results for singing voice separation.
The low vocal SDR means and high medians for all models again demonstrate the outlier problem discussed in Section~\ref{sec:results_quantitative_issues}.
The difference output layer does not noticeably change performance, as model M2 appears to be only very slightly better than model M1.
Initial experiments without fine-tuning showed a larger difference, which may indicate that a finer adjustment of weights makes constrained outputs less important, but they could still enable the usage of faster learning rates.
Introducing context noticeably improves performance, as model M3 shows, likely due to better predictions at output borders.
The stereo modeling in model M4 yields improvements especially for accompaniment, which may be because its sounds are panned more to the left or right channels than vocals.
The learned upsampling (M5) slightly improves the median, but slightly decreases the mean vocal SDR.
The small differences could be explained by the low number of weights in learned upsampling layers, considering that we also experimented with unconstrained convolutions, which brought more improvements but also high-frequency sound artifacts. We therefore consider M4 as our best model.
For multi-instrument separation, we achieve slightly lower but moderate performance (M6), as shown in Table~\ref{tab:multi_instrument_results}, in part due to less training data.

U7 performs worse than our comparison model M7, suggesting that our network architecture compares favourably to the state-of-the-art architecture since all else is kept constant during the experiments.
However, U7 stopped improving on the training set unexpectedly early, perhaps because it was not designed for minimising an audio-based MSE loss or because of effects related to backpropagating gradients through the inverse STFT. 
In contrast, U7a showed expected training behaviour using the magnitude-based loss. 
Our model also outperforms U7a, yielding considerably higher mean and median SDR scores.
The mean vocal SDR is the only exception, arising since our model has more outlier segments, but better output the majority of the time.

Models M4 and M6 were submitted as STL1 and STL2 to the SiSec campaign~\cite{Stoter2018}.
For vocals, M4 performs better or as well as almost all other systems.
Although it is significantly outperformed by submissions UHL3, TAK1-3 and TAU1, all of these except TAK1 used an additional 800 songs for training and thus have a large advantage.
M4 also separates accompaniment well, although slightly less so than the vocals. We
refer to~\cite{Stoter2018} for more details.

\begin{table}[t]
\begingroup
\scriptsize
\setlength{\tabcolsep}{3.8pt} 
\renewcommand{\arraystretch}{0.95} 
\begin{tabular}{|rr|rrrrr|rrr|}
\hline
& & M1 & M2 & M3 & M4 & M5 & M7 & U7 & U7a \\
\hline
\multirow{4}{*}{Voc.} & Med. & 3.90 & 3.92 & 3.96 & 4.46 &\textbf{4.58} & \textbf{3.49} & 2.76 & 2.74 \\
& MAD & 3.04 & 3.01 & 3.00 & 3.21 & 3.28 & 2.71 & 2.46 & 2.54 \\
& Mean & -0.12 & 0.05 & 0.31 & \textbf{0.65} & 0.55 & -0.23 & -0.66 & \textbf{0.51} \\
& SD & 14.00 & 13.63 & 13.25 & 13.67 & 13.84 & 13.00 & 12.38 & 10.82 \\
\hline
\multirow{4}{*}{Acc.} & Med. & 7.45 & 7.46 & 7.53 & \textbf{10.69} & 10.66 & \textbf{7.12} & 6.76 & 6.68 \\
& MAD & 2.08 & 2.10 & 2.11 & 3.15 & 3.10 & 2.04 & 2.00 & 2.04 \\
& Mean & 7.62 & 7.68 & 7.66 & \textbf{11.85} & 11.74 & \textbf{7.15} & 6.90 & 6.85 \\
& SD & 3.93 & 3.84 & 3.90 & 7.03 & 7.05 & 4.10 & 3.67 & 3.60 \\
\hline
\end{tabular}
\endgroup
\caption{Test set performance metrics (SDR statistics, in dB) for each singing voice separation model. Best performances overall and among comparison models are shown in bold.}
\label{tab:results}
\end{table}

\begin{table}[t]
\begingroup
\scriptsize
\setlength{\tabcolsep}{5.8pt} 
\begin{tabular}{|r|rrrr|rrrr|}
\hline
& \multicolumn{4}{c}{Vocals} & \multicolumn{4}{c|}{Other} \\
& Med. & MAD & Mean & SD & Med. & MAD & Mean & SD  \\
\hline
M6 & 3.0 & 2.76 & -2.10 & 15.41 & 2.03 & 1.64 & 1.68 & 6.14 \\
\hline
\hline
& \multicolumn{4}{c}{Bass} & \multicolumn{4}{c|}{Drums} \\
& Med. & MAD & Mean & SD & Med. & MAD & Mean & SD \\
\hline
M6 & 2.91 & 2.47 & -0.30 & 13.50 & 4.15 & 1.99 & 2.88 & 7.68 \\
\hline
\end{tabular}
\endgroup
\caption{Test performance metrics (SDR statistics, in dB) for our multi-instrument model}
\label{tab:multi_instrument_results}
\end{table}

\subsection{Qualitative results and observations}
\label{sec:results_qualitative}

\begin{figure}[t]
\vspace{-0.2cm}
\centering
\centerline{\includegraphics[width=8.5cm]{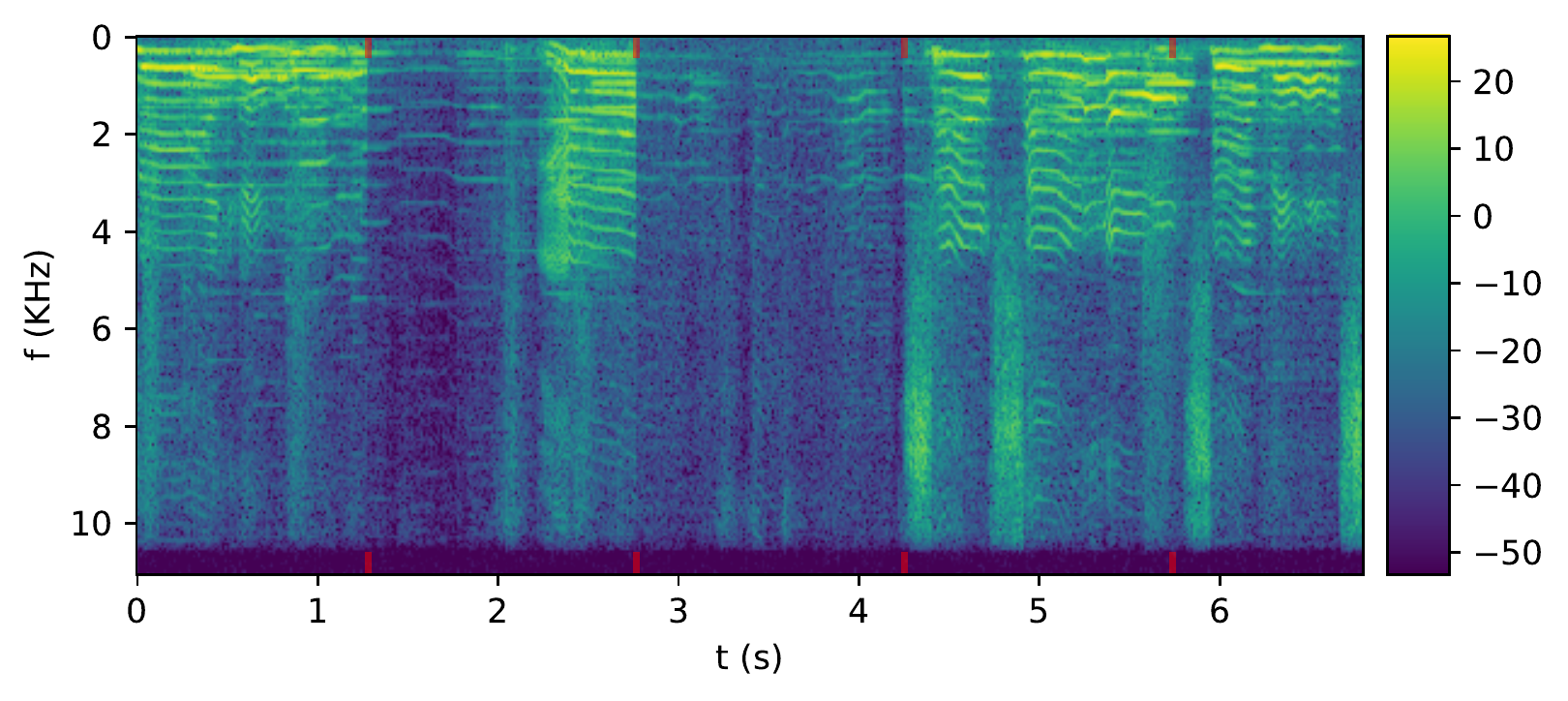}}
\vspace{-0.3cm}
\caption{Power spectrogram (dB) of a vocal estimate excerpt generated by a model without additional input context. Red markers show boundaries between independent segment-wise predictions.}
\label{fig:spectrogram}
\end{figure}

As an example of problems occurring when not using a proper temporal context, we generated a vocal source estimate for a song with the baseline model M1, and visualised an excerpt using a spectrogram in Figure~\ref{fig:spectrogram}.
Since the model's input and output are of equal length and the total output is created by concatenating predictions for non-overlapping consecutive audio segments, inconsistencies emerge at the borders shown in red: the loudness abruptly decreases at $1.2$ seconds, and a beginning vocal melisma is suddenly cut off at $2.8$ seconds, leaving only quiet noise, before the vocals reappear at $4.2$ seconds. A vocal melisma with only the vowel ``a" can sound similar to a non-vocal instrument and presumably was mistaken for one because no further temporal context was available.

In conclusion, these models suffer not only from inconsistencies at such segment borders, but are also less capable of performing separation there whenever information from a temporal context is required.
Larger input and output sizes alleviate the issue somewhat, but the problems at the borders remain. 
Blending the predictions for overlapping segments~\cite{Grais2018} is an ad-hoc solution, since the average of multiple predicted audio signals might not be a realistic prediction itself. For example, two sinusoids with equal amplitude and frequency, but opposite phase would cancel each other out.
Blending should thus be avoided in favour of our context-aware prediction framework.

\section{Discussion and conclusion}
\label{sec:conclusion}

In this paper, we proposed the Wave-U-Net for end-to-end audio source separation without any pre- or postprocessing, and applied it to singing voice and multi-instrument separation.
A long temporal context is processed by repeated downsampling and convolution of feature maps to combine high- and low-level features at different time-scales.
As indicated by our experiments, it outperforms the state-of-the-art spectrogram-based U-Net architecture~\cite{Jansson2017} when trained under comparable settings.
Since our data is quite limited in size however, it would be interesting to train our model on datasets comparable in size to the one used in~\cite{Jansson2017} to better assess respective advantages and disadvantages.

We highlight the lack of a proper temporal input context in recent separation and enhancement models, which can hurt performance and create artifacts, and propose a simple change to the padding of convolutions as a solution.
Similarly, artifacts resulting from upsampling by zero-padding as part of strided transposed convolutions can be addressed with a linear upsampling with a fixed or learned weight to avoid high-frequency artifacts.

Finally, we identify a problem in current SDR-based evaluation frameworks that produces outliers for quiet parts of sources and propose additionally reporting rank-based metrics as a simple workaround.
However, the underlying problem of perceptual evaluation of sound separation results using SDR metrics still remains and should be tackled at its root in the future.

For future work, we could investigate to which extent our model performs a spectral analysis, and how to incorporate computations similar to those in a multi-scale filterbank, or to explicitly compute a decomposition of the input signal into a hierarchical set of basis signals and weightings on which to perform the separation, similar to the TasNet~\cite{Luo2017a}.
Furthermore, better loss functions for raw audio prediction should be investigated such as the ones provided by generative adversarial networks~\cite{Goodfellow2014,Stoller2018}, since the MSE might not reflect the perceived loss of quality well.

\bibliography{refs}

%

\end{document}